\documentstyle[preprint,aps]{revtex}
\tighten
\draft
\widetext
\preprint{YITP-00-59}
\bigskip
\bigskip
\begin{document}
\title{Solitonic Brane World with Completely Localized (Super)Gravity}
\medskip
\author{Alberto Iglesias\footnote{E-mail:
iglesias@insti.physics.sunysb.edu} and Zurab Kakushadze\footnote{E-mail: 
zurab@insti.physics.sunysb.edu}}
\bigskip
\address{C.N. Yang Institute for Theoretical Physics\\ 
State University of New York, Stony Brook, NY 11794}

\date{November 12, 2000}
\bigskip
\medskip
\maketitle

\begin{abstract} 
{}We construct a solitonic 3-brane solution in the 5-dimensional 
Einstein-Hilbert-Gauss-Bonnet theory. This solitonic brane
is $\delta$-function like, and has the property that gravity is completely
localized on the brane. That is, there are no propagating degrees of freedom
in the bulk, while on the brane we have purely 4-dimensional Einstein gravity.
Thus, albeit the classical background is 5-dimensional, 
the quantum theory (perturbatively) is 4-dimensional. Our solution can 
be embedded in the supergravity context, where we have
completely localized supergravity 
on the corresponding solitonic brane, which is
a BPS object preserving 1/2 of the original supersymmetries. 
By including a scalar field, 
we also construct a smooth domain wall solution, 
which in a certain limit reduces to the 
$\delta$-function-like solitonic brane solution (this is possible
for the latter breaks diffeomorphisms only
spontaneously). We then show that in the smooth domain wall background the only
normalizable mode is the 4-dimensional graviton zero mode, while all 
the other (including massive
Kaluza-Klein) modes are not even plane-wave normalizable.
Finally, we observe that in 
compactifications of Type IIB on 5-dimensional Einstein manifolds other than
a 5-sphere the corresponding dual gauge theories on D3-branes are not conformal
in the ultra-violet, and at the quantum level
we expect the Einstein-Hilbert term to be generated 
in their world-volumes. We conjecture that in full 
string theory on Type IIB side this is due to higher curvature terms, which
cannot be ignored in such backgrounds. A stronger version of 
this conjecture also states that (at least in some cases)
in such backgrounds D3-branes are
solitonic objects with 
completely localized (super)gravity in their world-volumes.  
\end{abstract}
\pacs{}

\section{Introduction}

{}In the Brane World scenario the Standard Model gauge and matter fields
are assumed to be localized on  
branes (or an intersection thereof), while gravity lives in a larger
dimensional bulk of space-time 
\cite{early,BK,polchi,witt,lyk,shif,TeV,dienes,3gen,anto,ST,BW,Gog,RS,DGP,DG}. 
There is a big difference between the footings on which
gauge plus matter fields and gravity come in this picture\footnote{This, at
least in some sense, might not be an unwelcome feature - see, {\em e.g.},
\cite{witt,TeV,BW}.}. Thus, for instance, if gauge and matter fields are
localized on D-branes \cite{polchi}, they propagate only in the directions 
along the D-brane world-volume. Gravity, however, is generically 
not confined to the branes - even if we have a graviton zero mode localized
on the brane as in \cite{RS}, massive graviton modes are still free to 
propagate in the bulk.

{}In this paper we would like to ask the following question. Can we have
{\em complete} localization of gravity? As we will argue in the following,
the answer to this question appears to be positive. In particular, in a
certain setup, which we will describe in a moment, we will construct a
(flat) {\em solitonic} codimension-one brane world solution,
where gravity is 
completely localized on this solitonic brane. That is, the graviton
propagator in the bulk {\em vanishes}, while it is non-trivial on the
brane. In fact, in this solution the gravitational part of the brane
world-volume action is given by the usual 4-dimensional Einstein-Hilbert
term (assuming that the solitonic brane is a 3-brane). Moreover, even 
though we have a 5-dimensional theory (in particular, the classical
solitonic background is 5-dimensional), the quantum theory (at least 
perturbatively) is actually 4-dimensional. This is due to the fact that
in this solution there are no propagating bulk degrees of freedom,
so that there are no loop corrections in the bulk.

{}The setup within which we construct this solitonic brane world solution
is the 5-dimensional Einstein-Hilbert theory with a (negative) cosmological
term augmented with a Gauss-Bonnet coupling. The solitonic brane world
solution arises in this theory for a special value of the Gauss-Bonnet
coupling. The fact that there are no propagating degrees of freedom in the
bulk is then due to a perfect cancellation between the 
corresponding contributions coming from the Einstein-Hilbert and
Gauss-Bonnet terms, which occurs precisely for the value of the
Gauss-Bonnet coupling for which we have a solitonic brane world solution.
Since the bulk theory does not receive loop corrections, this setup
(at least perturbatively) is stable at the quantum level
as far as 5-dimensional physics is concerned. In particular, the
classical choice of parameters such as the special value of the
Gauss-Bonnet coupling (or the Gauss-Bonnet combination itself) does {\em not}
require order-by-order fine-tuning.

{}Since we essentially have a four-dimensional quantum theory, 
without supersymmetry generically we do expect a quantum instability
related to the 4-dimensional cosmological constant. In fact, our solitonic
brane world solution does admit curved deformations. However, if we embed
this solution in the (minimally) supersymmetric setup, then only the flat
solution does not break all the supersymmetries. In fact, the flat
solitonic brane world solution is a BPS state which preserves 1/2 of the
original supersymmetries. Moreover, we still have no propagating degrees of
freedom in the bulk, while on the brane we have completely localized 
{\em supergravity}. 

{}In the aforementioned solitonic brane world solution the brane is 
$\delta$-function like. Albeit seemingly strange, this is perfectly
consistent as this soliton does not break diffeomorphisms explicitly but
{\em spontaneously}. We should therefore be able to obtain this solitonic
brane world solution as a limit of a smooth solitonic brane world. We
show that this is indeed the case. We consider the system of 
Einstein-Hilbert-Gauss-Bonnet gravity coupled to a single scalar field with
a non-trivial scalar potential. For a suitable choice of the scalar
potential this system possesses kink-like solitonic domain wall solutions,
which break diffeomorphisms spontaneously. In a certain limit such a domain
wall solution gives precisely the aforementioned solitonic brane world
solution. We also point out that in this context the aforementioned
special choice of the Gauss-Bonnet coupling required for the solitonic
brane world solution to exist is essentially translated into the requirement
that the corresponding smooth domain wall solution have 4-dimensional
Poincar{\'e} invariance.   

{}The aforementioned smooth domain wall solution has the property that 
only the 4-dimensional graviton zero mode is normalizable in this background
(this zero mode is quadratically normalizable). In particular, none of the
massive modes are even plane-wave normalizable in this background. This 
makes it clear why in the limit, where we recover the $\delta$-function-like
solitonic brane world solution, gravity is completely localized on the brane.
On the other hand, the fact that only the graviton zero mode is normalizable
in the smooth domain wall background would not be possible without the higher
curvature terms. This indicates that inclusion of higher curvature terms
can in some cases qualitatively change gravity in brane world scenarios. 

{}At the end of the paper we speculate on a possible realization of our
scenario within string theory. In particular, we observe that in 
compactifications of Type IIB string theory on $X_5$, where $X_5$ is a 
5-dimensional Einstein manifold, the dual field theory
generically is conformal only in the infra-red but not 
in the ultra-violet (in the special case where $X_5$ is a 5-sphere, we have
the ${\cal N}=4$ SYM theory which is scale invariant). In such cases we
therefore expect that quantum corrections will generically 
generate \cite{DGP,DG} (among other terms) the Einstein-Hilbert term in the
world-volume of the corresponding D3-branes. We conjecture that in the dual
Type IIB picture the appearance of this term is due to {\em higher curvature}
terms which should be present in such backgrounds (which have reduced
number of unbroken supersymmetries). That is, we suggest that higher
curvature terms are important in such backgrounds, and should not be ignored.
We also propose a stronger version of
this conjecture according to which in full string theory D3-branes with
such non-conformal theories in their world-volumes (at least in some cases)
are solitonic objects similar to the solitonic brane world we discuss in this 
paper, and we have completely localized (super)gravity (plus
(super)Yang-Mills) on the branes. If this conjecture indeed holds, 
our scenario
might have interesting phenomenological implications such as at least 
a partial solution to the moduli problem in string compactifications, as
well as a possibility of 
having truly 4-dimensional gravity localized on a brane in
non-compact extra space.

\section{The Setup}

{}In this section we discuss the setup within which we will discuss the
aforementioned solitonic brane world solution. The action for this model
is given by (for calculational convenience we will keep the number of 
space-time dimensions $D$ unspecified):
\begin{equation}\label{actionGB}
 S=M_P^{D-2}
 \int d^D x \sqrt{-G} \left\{R+\lambda\left[R^2-4R_{MN}^2+R_{MNST}^2\right]
 -\Lambda\right\}~,
\end{equation}
where $M_P$ is the $D$-dimensional (reduced) Planck scale, and 
the Gauss-Bonnet coupling $\lambda$ has dimension $({\rm length})^2$.
Finally, the bulk vacuum energy density $\Lambda$ is a constant.

{}The equations of motion following form the action (\ref{actionGB}) read:
\begin{eqnarray}
 &&R_{MN}-{1\over 2}G_{MN} R
 -{1\over 2}\lambda G_{MN}\left(R^2-4R^{MN}R_{MN}+R^{MNRS}R_{MNRS}
 \right)+\nonumber\\
 \label{einstein}
 &&2\lambda\left(R R_{MN}-2R_{MS}{R^S}_N+R_{MRST}{R_N}^{RST}-2R^{RS}R_{MRNS}
 \right)+{1\over 2}G_{MN} \Lambda=0~.
\end{eqnarray}
Note that this equation does not contain terms with third and fourth 
derivatives of the metric.

{}In the following we will be interested in solutions 
to the above equations of motion 
with the warped \cite{Visser} metric of the form 
\begin{equation}\label{warped}
 ds_D^2=\exp(2A)\eta_{MN} dx^M dx^N~,
\end{equation}
where $\eta_{MN}$ is the flat $D$-dimensional Minkowski metric, and
the warp factor $A$, which is a function of $z\equiv x^D$, 
is independent of the other $(D-1)$ coordinates $x^\mu$.
With this ans{\"a}tz, we have the following
equations of motion for $A$ (prime denotes derivative w.r.t. $z$):
\begin{eqnarray}
 \label{A'GB}
 &&(D-1)(D-2)(A^\prime)^2\left[1-(D-3)(D-4)\lambda(A^\prime)^2
 \exp(-2A)\right]+ \Lambda \exp(2A)=0~,\\
 \label{A''GB}
 &&(D-2)\left[A^{\prime\prime}-(A^\prime)^2\right]
 \left[1-2(D-3)(D-4)\lambda(A^\prime)^2
 \exp(-2A)\right]
 =0~.
\end{eqnarray}
This system of equations has a set of solutions where the $D$-dimensional
space is an AdS space for a continuous range of parameters $\Lambda$ and 
$\lambda$. The volume of the $z$ direction for this set of solutions is
infinite.

{}There, however, also exists a solution where the volume of the
$z$ direction is finite if we ``fine-tune'' the Gauss-Bonnet coupling 
$\lambda$ and the bulk vacuum energy density $\Lambda$ as 
follows\footnote{This special value of the Gauss-Bonnet coupling has appeared
in a somewhat different context in \cite{Zanelli}.}:
\begin{equation}\label{fine}
 \Lambda=- {{(D-1)(D-2)}\over{(D-3)(D-4)}}{1\over 4\lambda}~,
\end{equation}
where $\lambda>0$, and $\Lambda<0$. 
This solution is given by (we have chosen the 
integration constant such that $A(0)=0$):
\begin{equation}\label{solutionGB}
 A(z)=-\ln\left[{|z|\over\Delta}+1\right]~,
\end{equation}
where $\Delta$ is given by
\begin{equation}\label{Delta}
 \Delta^2= 2(D-3)(D-4)\lambda~.
\end{equation}
Note that $\Delta$ can be positive or negative. In the former case the
volume of the $z$ direction is finite: $v=2\Delta/(D-1)$. 
On the other hand, in the latter case it is infinite.
As we will see in the following, the negative $\Delta$ case corresponds to
a non-unitary theory. 

{}Note that $A^\prime$ is discontinuous at $z=0$, and $A^{\prime\prime}$
has a $\delta$-function-like behavior at $z=0$. Note, however, that
(\ref{A''GB}) is still satisfied as in this solution
\begin{equation}\label{factor}
 1-2(D-3)(D-4)\lambda(A^\prime)^2\exp(-2A)=0~.
\end{equation}
Thus, this solution
describes a codimension one soliton. The tension of this soliton,
which is given by
\begin{equation}
 f={4(D-2)\over \Delta}M_P^{D-2}~,
\end{equation}
is positive for $\Delta>0$, and it is negative for $\Delta<0$. The
aforementioned non-unitarity in the latter case is, in fact, attributed to
the negativity of the brane tension. Here and in the following we refer to the
$z=0$ hypersurface, call it $\Sigma$, as the brane.

\section{Gravity in the Solitonic Brane World}

{}In this section we would like to study gravity in the solitonic brane
world solution discussed in the previous section.
To do this, let us study small fluctuations around the solution:
\begin{equation}\label{fluctu}
 G_{MN}=\exp(2A)\left[\eta_{MN}+{\widetilde h}_{MN}\right]~,
\end{equation}
where for convenience reasons we have chosen to work with 
${\widetilde h}_{MN}$
instead of metric fluctuations $h_{MN}=\exp(2A){\widetilde h}_{MN}$. 

{}To proceed further, we need equations of motion for ${\widetilde h}_{MN}$. 
Let us assume that we have matter localized on the brane, and 
let the corresponding conserved energy-momentum tensor be $T_{\mu\nu}$:
\begin{equation}\label{conserved}
 \partial^\mu T_{\mu\nu}=0~.
\end{equation}
The graviton field ${\widetilde h}_{\mu\nu}$ couples to $T_{\mu\nu}$ via
the following term in the action (note that ${\widetilde h}_{\mu\nu}=
h_{\mu\nu}$ at $z=0$ as we have set $A(0)=0$):
\begin{equation}\label{int}
 S_{\rm {\small int}}={1\over 2} \int_\Sigma d^{D-1} x ~T_{\mu\nu}
 {\widetilde h}^{\mu\nu}~.
\end{equation} 
In the following we will use the following notations for the component
fields:
\begin{equation}
 H_{\mu\nu}\equiv{\widetilde h}_{\mu\nu}~,~~~A_\mu\equiv
 {\widetilde h}_{\mu D}~,~~~\rho\equiv{\widetilde h}_{DD}~.
\end{equation}
 The linearized equations of motion for the component fields $H_{\mu\nu}$,
$A_\mu$ and $\rho$ read:
\begin{eqnarray}\label{EOM1oGB}
 &&\left[1-2(D-3)(D-4)\lambda(A^\prime)^2\exp(-2A)\right]\Big(
 \partial_\sigma\partial^\sigma 
 H_{\mu\nu} +\partial_\mu\partial_\nu
 H-\partial_\mu \partial^\sigma H_{\sigma\nu}-
 \partial_\nu \partial^\sigma H_{\sigma\mu}-\nonumber\\
 &&\eta_{\mu\nu}
 \left[\partial_\sigma\partial^\sigma H-\partial^\sigma\partial^\rho
 H_{\sigma\rho}\right]+
 H_{\mu\nu}^{\prime\prime}-\eta_{\mu\nu}H^{\prime\prime}+
 (D-2)A^\prime\left[H_{\mu\nu}^\prime-\eta_{\mu\nu}H^\prime\right]-
 \nonumber\\
 &&\left\{\partial_\mu A_\nu^\prime + \partial_\nu A_\mu^\prime -
 2\eta_{\mu\nu}\partial^\sigma A_\sigma^\prime+ (D-2)A^\prime
 \left[\partial_\mu A_\nu +\partial_\nu A_\mu
 -2\eta_{\mu\nu}\partial^\sigma
 A_\sigma\right]\right\}\nonumber+\\
 &&\left\{\partial_\mu\partial_\nu\rho-\eta_{\mu\nu}
 \partial_\sigma\partial^\sigma 
 \rho+\eta_{\mu\nu}\left[(D-2)A^\prime\rho^\prime
 +(D-1)(D-2)(A^\prime)^2\rho\right]\right\}\Big)-\nonumber\\
 &&4(D-4)\lambda\left[A^{\prime\prime}-(A^\prime)^2\right]\exp(-2A)
 \Big(\partial_\sigma\partial^\sigma 
 H_{\mu\nu} +\partial_\mu\partial_\nu
 H-\partial_\mu \partial^\sigma H_{\sigma\nu}-
 \partial_\nu \partial^\sigma H_{\sigma\mu}-\nonumber\\
 &&\eta_{\mu\nu}
 \left[\partial_\sigma\partial^\sigma H-\partial^\sigma\partial^\rho
 H_{\sigma\rho}\right]+
 (D-3)A^\prime\left\{H_{\mu\nu}^\prime-\partial_\mu A_\nu-
 \partial_\nu A_\mu -\eta_{\mu\nu}\left[H^\prime-2\partial^\sigma A_\sigma
 \right]\right\} \Big)+\nonumber\\
 && 2(D-2)\left[A^{\prime\prime}-(A^\prime)^2\right]
 \left[1-4(D-3)(D-4)\lambda(A^\prime)^2\exp(-2A)\right]\eta_{\mu\nu}\rho=
 \nonumber\\
 &&-M_P^{2-D} T_{\mu\nu} \delta(z)~,\\ 
 \label{EOM2oGB} 
 &&\left[1-2(D-3)(D-4)\lambda(A^\prime)^2\exp(-2A)\right]\Big(
 \left[\partial^\mu H_{\mu\nu}-\partial_\nu H\right]^\prime 
 -\partial^\mu F_{\mu\nu}+\nonumber\\
 && (D-2)A^\prime \partial_\nu\rho\Big)=0~,\\
 \label{EOM3oGB}
 &&\left[1-2(D-3)(D-4)\lambda(A^\prime)^2\exp(-2A)\right]
 \Big(-\left[\partial^\mu\partial^\nu H_{\mu\nu}-\partial^\mu\partial_\mu H
 \right] +\nonumber\\
 &&(D-2) A^\prime \left[H^\prime-2\partial^\sigma A_\sigma\right]
 -(D-1)(D-2)(A^\prime)^2\rho\Big)=0~,
\end{eqnarray}
where $F_{\mu\nu}\equiv\partial_\mu A_\nu - \partial_\nu A_\mu$ 
is the $U(1)$ field strength for the graviphoton, and $H\equiv {H_\mu}^\mu$.

{}The above equations of motion are invariant under certain gauge
transformations corresponding to unbroken diffeomorphisms.  
In terms of ${\widetilde h}_{MN}$ the full
$D$-dimensional diffeomorphisms 
\begin{equation}
 \delta h_{MN}=\nabla_M\xi_N+\nabla_N\xi_M
\end{equation}
are given by the following gauge 
transformations (here we use $\xi_M\equiv \exp(2A){\widetilde \xi}_M$):
\begin{equation}\label{gauge}
 \delta{\widetilde h}_{MN}=\partial_M {\widetilde\xi}_N+
 \partial_N{\widetilde\xi}_M+2A^\prime\eta_{MN} \omega~, 
\end{equation} 
where $\omega\equiv {\widetilde \xi}_D$.
In terms of the component fields $H_{\mu\nu}$, $A_\mu$ and $\rho$, the
full $D$-dimensional diffeomorphisms read:
\begin{eqnarray}\label{diff1}
 &&\delta H_{\mu\nu}=\partial_\mu{\widetilde\xi}_\nu+\partial_\nu{\widetilde
 \xi}_\mu+2\eta_{\mu\nu}A^\prime\omega~,\\
 \label{diff2}
 &&\delta A_\mu={\widetilde\xi_\mu}^\prime +\partial_\mu \omega~,\\
 \label{diff3}
 &&\delta\rho=2\omega^\prime+2A^\prime\omega~.
\end{eqnarray}
It is not difficult to check that
the equations of motion (\ref{EOM1oGB}), (\ref{EOM2oGB}) and (\ref{EOM3oGB})
are invariant under these full $D$-dimensional diffeomorphisms. That is, there
are no restrictions on $\omega$ or ${\widetilde \xi}_\mu$ or derivatives 
thereof including at $z=0$. In particular, this is the case for the solitonic
brane world solution despite its $\delta$-function-like structure. The reason 
for this is that this solution being a soliton 
does not break the full $D$-dimensional 
diffeomorphisms explicitly but only {\em spontaneously}.

{}Since we have the full $D$-dimensional diffeomorphisms, we can always gauge
$A_\mu$ and $\rho$ away. In fact, in the following we will see that for the
solitonic brane world background this can indeed be done without introducing 
any inconsistencies. However, before we adapt this gauge fixing and solve 
the above equations of motion, we would like to make the following 
important observation. Note that for the solitonic brane world solution
(\ref{solutionGB}) with $\Delta$ given by (\ref{Delta}) we have (\ref{factor}).
On the other hand, this vanishing factor is precisely the one that multiplies
the terms in (\ref{EOM1oGB}), (\ref{EOM2oGB}) and (\ref{EOM3oGB}) 
corresponding to the propagation of the fields $H_{\mu\nu}$, $A_\mu$ and
$\rho$ in the bulk. That is, in the solitonic brane world solution these fields
do not propagate in the $z$ direction at all. This is due to a cancellation
between contributions of the Einstein-Hilbert and Gauss-Bonnet terms into
the bulk propagator in this background\footnote{Note that in 
(\ref{EOM1oGB}), (\ref{EOM2oGB}) and (\ref{EOM3oGB}) the terms
that survive in the limit where the warp factor $A$ is a constant 
correspond to the terms that arise upon linearization of the
$D$-dimensional Einstein-Hilbert term around the {\em flat} background.
On the other hand, there are no such terms corresponding to 
linearization of the
$D$-dimensional Gauss-Bonnet term around the flat background. This is due to
the fact that even in $D>4$ the terms quadratic in metric fluctuations 
coming from expanding the Gauss-Bonnet term around the flat background give
rise to a total derivative in the action \cite{Zwiebach,Zumino} (in $D=4$
the Gauss-Bonnet term is a total derivative altogether as it corresponds to
the 4-dimensional Euler invariant).}.
On the other hand, (some of)
these fields do propagate on the brane. Indeed, in the above background we
have 
\begin{equation}
 A^{\prime\prime}-(A^\prime)^2=-{2\over \Delta} \delta(z)~.
\end{equation}
Then (\ref{EOM1oGB}) gives the following equation of motion (note that
(\ref{EOM2oGB}) and (\ref{EOM3oGB}) are trivially satisfied in this 
background):
\begin{eqnarray}\label{EOM1oGBX}
 &&\Big(\partial_\sigma\partial^\sigma 
 H_{\mu\nu} +\partial_\mu\partial_\nu
 H-\partial_\mu \partial^\sigma H_{\sigma\nu}-
 \partial_\nu \partial^\sigma H_{\sigma\mu}-\eta_{\mu\nu}
 \left[\partial_\sigma\partial^\sigma H-\partial^\sigma\partial^\rho
 H_{\sigma\rho}\right]+\nonumber\\
 &&(D-3)A^\prime\left\{H_{\mu\nu}^\prime-\partial_\mu A_\nu-
 \partial_\nu A_\mu -\eta_{\mu\nu}\left[H^\prime-2\partial^\sigma A_\sigma
 \right]\right\}+
 (D-2)(D-3)\Delta^{-2}\eta_{\mu\nu}\rho\Big)\delta(z)=\nonumber\\
 &&-{\widehat M}_P^{3-D} 
 T_{\mu\nu} \delta(z)~,
\end{eqnarray}
where
\begin{equation}
 {\widehat M}_P^{D-3}\equiv {4\Delta\over{D-3}}M_P^{D-2}~,
\end{equation}
and in the following we will identify ${\widehat M}_P$ with the 
$(D-1)$-dimensional Planck scale.

\subsection{Completely Localized Gravity}

{}Next, we would like to see what is the solution to the equation of
motion (\ref{EOM1oGBX}). First, note that, as we have already mentioned,
we can always gauge $A_\mu$ and $\rho$ away. That is, these fields are {\em
not} propagating degrees of freedom. Note that after this gauge fixing the
residual gauge symmetry is given by the $(D-1)$-dimensional diffeomorphisms
for which $\omega\equiv 0$, and ${\widetilde \xi}_\mu$ are independent of $z$.
Second, note that the term in the curly
brackets in (\ref{EOM1oGBX}) is multiplied by
$A^\prime\delta(z)$. This quantity, 
however, is vanishing as $A^\prime$ has a ${\rm sign}(z)$-like discontinuity
at $z=0$. We therefore obtain the following equation of motion for the
$(D-1)$-dimensional graviton components $H_{\mu\nu}$:
\begin{equation}\label{EOM1oGBXY}
 \Big(\partial_\sigma\partial^\sigma 
 H_{\mu\nu} +\partial_\mu\partial_\nu
 H-\partial_\mu \partial^\sigma H_{\sigma\nu}-
 \partial_\nu \partial^\sigma H_{\sigma\mu}-\eta_{\mu\nu}
 \left[\partial_\sigma\partial^\sigma H-\partial^\sigma\partial^\rho
 H_{\sigma\rho}\right]+{\widehat M}_P^{3-D} 
 T_{\mu\nu}\Big) \delta(z)=0~.
\end{equation}
Note that this equation is purely $(D-1)$-dimensional. Thus, gravity is
{\em  completely} localized on the brane, that is at the $z=0$ hypersurface
$\Sigma$. In particular, the graviton field $H_{\mu\nu}$
is non-vanishing only on the brane, while it vanishes in the bulk:
\begin{equation}\label{Hbulk}
 H_{\mu\nu}(z\not=0)=0~.
\end{equation}
Note that (\ref{EOM1oGBXY}) does not by itself imply 
(\ref{Hbulk}). In particular, {\em a priori} $H_{\mu\nu}$ at $z\not=0$ can be
arbitrary. However, as we explained above, we have no propagating degrees of
freedom in the bulk, that is, the graviton propagator in the bulk vanishes, 
while it is non-vanishing only on the brane. This implies that perturbations
due to matter localized on the brane should {\em not} propagate into the bulk
but only on the brane, hence (\ref{Hbulk}).

{}On the brane (\ref{EOM1oGBXY}) can be solved in a standard way. Thus, in the
harmonic gauge (we can use this or any other suitable gauge fixing on the
brane as we have unbroken $(D-1)$-dimensional diffeomorphisms)
\begin{equation}
 \partial^\mu H_{\mu\nu}={1\over 2}\partial_\nu H
\end{equation}
we have
\begin{equation}\label{4Dgrav}
 H_{\mu\nu}(p,z=0)={\widehat M}_P^{3-D}{1\over p^2}\left[T_{\mu\nu}(p)-
 {1\over{D-3}}\eta_{\mu\nu} T(p)\right]~,
\end{equation}
where we have performed the Fourier transform w.r.t. the $(D-1)$-dimensional 
coordinates $x^\mu$ (the corresponding momenta are $p^\mu$, and $p^2\equiv
p^\mu p_\mu$), and $T(p)\equiv {T_\mu}^\mu(p)$. From (\ref{EOM1oGBXY})
as well as (\ref{4Dgrav}) it is clear that ${\widehat M}_P$ is the 
$(D-1)$-dimensional Planck scale for $(D-1)$-dimensional 
gravity localized on the brane (note that the momentum and tensor
structures in (\ref{4Dgrav}) are $(D-1)$-dimensional). Actually, 
${\widehat M}_P$ is identified with the $(D-1)$-dimensional Planck scale
for the positive $\Delta$ solution. As to the negative $\Delta$ solution,  
we have ``antigravity'' localized on the brane, and the corresponding
theory is non-unitary due to negative norm states propagating on the brane.

{}Note that above our analysis was confined to the linearized theory.
The above conclusions, however, are valid in the full non-linear theory.
Indeed, we have no propagating degrees of freedom in the bulk, while on the
brane we have only the zero mode for the $(D-1)$-dimensional graviton 
components $H_{\mu\nu}$. This then implies that in the solitonic brane
world background (the gravitational part of) the
brane world-volume theory is described by the
$(D-1)$-dimensional Einstein-Hilbert action:
\begin{equation}\label{actionWV}
 S_{\rm {\small brane}}={\widehat M}_P^{D-2}
 \int d^{D-1} x \sqrt{-{\widehat G}} 
 {\widehat R}~,
\end{equation}
where ${\widehat G}_{\mu\nu}$ is the $(D-1)$-dimensional 
metric on the brane; all the
hatted quantities are $(D-1)$-dimensional, and are constructed from 
${\widehat G}_{\mu\nu}$. Note that there is no 
$(D-1)$-dimensional Gauss-Bonnet term in this action, which can be seen by
examining (\ref{einstein}).

\section{Quantum Stability}

{}The solitonic brane world solution we discussed in the previous sections
has the following remarkable property - the bulk theory does not receive
any loop corrections. Indeed, there are no propagating degrees of freedom
in the bulk, hence the absence of loop corrections\footnote{From now on,
when referring to quantum stability or absence of quantum corrections, we
mean perturbatively. {\em A priori} there might be non-perturbative
corrections in the bulk which might modify some of the following
conclusions, but this issue is outside of the scope of this paper.}. 
This implies that 
the bulk action is not renormalized at all, and, in
particular, the relation (\ref{fine}) between the bulk vacuum energy
density $\Lambda$ and the Gauss-Bonnet coupling $\lambda$ is 
stable against quantum corrections. This is why we used the word
``fine-tuning'' in quotation marks in section II - once we choose the
parameters of the classical theory to satisfy (\ref{fine}), 
we need no fine-tuning at the quantum level.

{}The above observation has important implications
for gravity in the solitonic brane world solution. First, 
there is no danger of delocalization of gravity, which is generically
expected to occur at the quantum level due to higher curvature bulk terms
\cite{COSM,zuraRS,olindo} in warped backgrounds such as \cite{RS}.     
Second, due to the spontaneous nature of diffeomorphism breaking in the
solitonic brane world, the graviscalar and graviphoton components are pure 
gauge degrees of freedom. This implies that at the quantum level 
there is no danger of
generating brane world-volume terms involving, say, the graviscalar, which 
are generically expected to lead to inconsistencies in the coupling
between bulk gravity and brane matter \cite{zuraRS} in warped backgrounds
such as \cite{RS}. {}These properties of the solitonic brane world can be
understood in a simple way by noting that the quantum 
theory is actually $(D-1)$-dimensional (albeit the classical background is
$D$-dimensional), so the only quantum instability we can expect is that 
related to the $(D-1)$-dimensional physics.

\subsection{Curved Deformations}

{}Such an instability generically indeed exists as we are dealing with a 
theory containing gravity - without supersymmetry we expect that
generically $(D-1)$-dimensional cosmological constant will be generated at
the quantum level. Here we would like to verify that the solitonic brane
world indeed admits curved deformations. 

{}Thus, instead of the flat ans{\"a}tz (\ref{warped}), 
let us look for solutions with the following warped metric 
\begin{equation}\label{warpedcurved}
 ds_D^2=\exp(2A)\left[{\widehat g}_{\mu\nu} dx^\mu dx^\nu +(dz)^2\right]~,
\end{equation}
where the $(D-1)$-dimensional background metric ${\widehat g}_{\mu\nu}$ is
independent of $z$, but need not be flat.
The equations of motion (\ref{einstein}) then give the
following equations of motion for $A$:
\begin{eqnarray}
 \label{A'curved}
 &&(D-1)(D-2)(A^\prime)^2\left[1-(D-3)(D-4)\lambda(A^\prime)^2\exp(-2A)\right]
 +\Lambda\exp(2A)-\nonumber\\
 &&{D-1\over D-3} {\widehat \Lambda}\left[1-2(D-3)(D-4)\lambda
 (A^\prime)^2\exp(-2A)\right]
 -\lambda{\widehat\chi}\exp(-2A)=0~,\\
 \label{A''curved}
 &&(D-2)\left[A^{\prime\prime}-(A^\prime)^2\right]
 \left[1-2(D-3)(D-4)\lambda(A^\prime)^2\exp(-2A)+2
 {D-4\over D-2}\lambda{\widehat\Lambda}
 \exp(-2A)\right]+\nonumber\\
 &&{1\over D-3}{\widehat \Lambda}
 \left[1-2(D-3)(D-4)\lambda(A^\prime)^2\exp(-2A)\right]+{2\lambda\over D-1}
 {\widehat\chi}\exp(-2A)=0~.
\end{eqnarray}
Here ${\widehat \Lambda}$ is the cosmological constant of the 
$(D-1)$-dimensional manifold, which is therefore an Einstein manifold, 
described by the metric ${\widehat g}_{\mu\nu}$. 
Our normalization of ${\widehat \Lambda}$ is such
that the $(D-1)$-dimensional metric ${\widehat g}_{\mu\nu}$ satisfies
Einstein's equations
\begin{equation}
 {\widehat R}_{\mu\nu}-{1\over 2}{\widehat g}_{\mu\nu}
 {\widehat R}=-{1\over 2}
{\widehat g}_{\mu\nu}{\widehat\Lambda}~.
\end{equation}
Moreover, the quantity 
\begin{equation}\label{chi}
 {\widehat\chi}\equiv{\widehat R}^2-4{\widehat R}_{\mu\nu}^2+
 {\widehat R}_{\mu\nu\sigma\tau}^2
\end{equation} 
is also a constant (for $\lambda\not=0$). 
Finally, the aforementioned
$(D-1)$-dimensional Einstein manifold must be such that
\begin{equation}\label{condi}
 {\widehat R}_{\mu\rho\sigma\tau}{{\widehat R}_\nu}^{\,\,\,\,\rho\sigma\tau}=
 {1\over D-1}{\widehat R}_{\lambda\rho\sigma\tau}^2 {\widehat g}_{\mu\nu}~.
\end{equation}
This condition is automatically satisfied for maximally symmetric Einstein 
manifolds. 

{}Albeit the Einstein manifold described by the metric ${\widehat
g}_{\mu\nu}$ {\em a priori}
need not be maximally symmetric, for our purposes here it will
suffice to confine our attention to maximally symmetric cases. Then we have 
\begin{equation}
 {\widehat R}_{\mu\nu\rho\sigma}={{\widehat\Lambda}\over(D-2)(D-3)}
 \left({\widehat g}_{\mu\rho}{\widehat g}_{\nu\sigma}-
 {\widehat g}_{\mu\sigma}{\widehat g}_{\nu\rho}\right)~,
\end{equation} 
and
\begin{equation}
 {\widehat\chi}={(D-1)(D-4)\over (D-2)(D-3)}{\widehat\Lambda}^2~. 
\end{equation}
It is then not difficult to see that, for the special value of the
bulk vacuum energy density given by (\ref{fine}), the second order equation
(\ref{A''curved}) is automatically satisfied as long as the first order 
equation (\ref{A'curved}) is satisfied. In fact, this latter equation
simplifies as follows:
\begin{equation}\label{simplecurved}
 (A^\prime)^2={{\widehat\Lambda}\over(D-2)(D-3)}+{1\over \Delta^2}\exp(2A)~,
\end{equation} 
where $\Delta$ is given by (\ref{Delta}). This equation has consistent
smooth solutions (that is, solutions with continuous derivatives of $A$)
for any ${\widehat\Lambda}$. These solutions correspond to foliations of
the $D$-dimensional AdS space with the vacuum energy density 
$\Lambda_1=2\Lambda$ with maximally symmetric $(D-1)$-dimensional Einstein
manifolds. On the other
hand, as in the flat case, we also have solitonic brane world solutions.
Thus, let $A_1 (z)$ be a smooth solution of the aforementioned type. Then
the solution
\begin{equation}\label{solitoncurved}
 A(z)=A_1(|z|)
\end{equation} 
describes a solitonic brane world solution. If at $z\rightarrow +\infty$ 
we have $A_1(z)\rightarrow -\infty$, then (\ref{solitoncurved}) gives a
curved deformation of a flat solution
with positive brane tension. If at $z\rightarrow +\infty$ 
we have $A_1(z)\rightarrow +\infty$, then (\ref{solitoncurved}) gives
a curved deformation of a flat solution
with negative brane tension. The former type of solutions give a consistent
curved solitonic brane world. 

{}Here the following remark is in order. The solitonic brane world
solutions obtained via the above procedure have discontinuous $A^\prime$
for generic values of ${\widehat\Lambda}$. There is, however, a special
value of ${\widehat\Lambda}$, given by
\begin{equation}\label{specialCC}
 {\widehat\Lambda}=-(D-2)(D-3)\Delta^{-2}~,
\end{equation} 
for which $A_1(z)$ is symmetric w.r.t. the reflection $z\rightarrow-z$,
so that $A(z)$ has a continuous derivative everywhere. That is, the
solitonic brane world solution given by (\ref{solitoncurved})
in this case coincides with the corresponding smooth solution. From now on,
when referring to the solitonic brane world solution, we will therefore
{\em not} include the case where ${\widehat \Lambda}$
satisfies (\ref{specialCC}).

\subsection{Supersymmetry and BPS Solitonic Brane World}

{}Thus, as expected, {\em a priori} we can have non-vanishing cosmological
constant on the above solitonic brane. However, if we embed this solution
in the supersymmetric setup, then only the flat solution does not break all
the supersymmetries\footnote{As we pointed out in the previous subsection,
if the cosmological constant on the brane satisfies (\ref{specialCC}),
then the corresponding solution is actually smooth. In this case all the
supersymmetries are therefore intact, but we do not interpret this solution
as the solitonic brane world solution.}. Here
we would like to discuss this supersymmetric generalization in more detail.

{}Note that the action (\ref{actionGB}) can be supersymmetrized (for
the standard values of $D$). Upon supersymmetrization, we have fields
other than the metric. Let us consider backgrounds where all bosonic
fields other than the metric have vanishing expectation values. Then the
solitonic brane world solution is the same as before. Since graviton does
not propagate in the bulk in this background, it then follows by
supersymmetry that no other fields will have propagating degrees of freedom
in the bulk either (here we are assuming that we have minimal 
supersymmetry). 
That is, the cancellation between the bulk Einstein-Hilbert
and Gauss-Bonnet terms that we saw above generalizes to all the
other (superpartner) terms in the bulk as well. 
On the other hand, on the brane
we do have propagating degrees of freedom. In fact, we expect to have 
completely localized {\em supergravity} on the brane if there are some unbroken
supersymmetries. Here we would like to show that ${\widehat\Lambda}\not=0$
solutions break all supersymmetries (regardless of the sign of
${\widehat\Lambda}$), while the flat solitonic brane world is a BPS
solution which preserves 1/2 of the original supersymmetries.

{}To see this, let us study Killing spinors in the solitonic
brane world background:
\begin{equation}
 {\cal D}_M \epsilon = 0~.
\end{equation}
Here ${\cal D}_M$ is a generalized covariant derivative:
\begin{equation}
 {\cal D}_M=D_M-{1\over 2} W\Gamma_M~,
\end{equation}
where $D_M$ is the usual covariant derivative containing the spin
connection, and $W$ is interpreted as
the superpotential. The $D$-dimensional gamma matrices $\Gamma_M$ are
defined via
\begin{equation}
 \left\{\Gamma_M~,~\Gamma_N\right\}=2g_{MN}~,
\end{equation}
where $g_{MN}=\exp(2A){\widetilde g}_{MN}$ is the background metric:
${\widetilde g}_{\mu\nu}={\widehat g}_{\mu\nu}$, ${\widetilde g}_{\mu D}=0$,
and ${\widetilde g}_{DD}=1$.

{}It is not difficult to show that in such warped backgrounds we have
\begin{eqnarray}\label{Killing1}
 && 0={\cal D}_D\epsilon=\epsilon^\prime-{1\over 2}W \exp(A){\widetilde 
 \Gamma}_D\epsilon~,\\
 \label{Killing2}
 && 0={\cal D}_\mu\epsilon={\widehat D}_\mu\epsilon +{1\over 2}
 {\widehat \Gamma}_\mu\left[A^\prime{\widetilde \Gamma}_D-W\exp(A)\right]
 \epsilon~,
\end{eqnarray} 
where ${\widehat D}_\mu$ is the $(D-1)$-dimensional covariant derivative 
corresponding to the metric ${\widehat g}_{\mu\nu}$, while 
${\widehat\Gamma}_\mu={\widetilde \Gamma}_\mu\equiv \exp(-A)\Gamma_\mu$ 
are the
$(D-1)$-dimensional gamma matrices corresponding to the metric 
${\widehat g}_{\mu\nu}$, and 
${\widetilde\Gamma}_D\equiv \exp(-A)\Gamma_D$ 
is a constant matrix.

{}Before solving the Killing spinor equations, let us note that to define an
unbroken supercharge for a given Killing spinor we must make sure that
the global integrability conditions
\begin{equation}
 \left[{\cal D}_M~,~{\cal D}_N\right]\epsilon=0
\end{equation}
are also satisfied. In the component form these conditions read:
\begin{eqnarray}\label{condi1}
 &&0=\left[{\cal D}_\mu~,~{\cal D}_\nu\right]\epsilon=
 {1\over 4}\left({1\over 2} {\widehat R}_{\mu\nu\sigma\tau}
 \left[{\widehat\Gamma}^\sigma~,~{\widehat\Gamma}^\tau\right]+
 \left[W^2\exp(2A)-(A^\prime)^2\right]
 \left[{\widehat\Gamma}_\mu~,~{\widehat\Gamma}_\nu\right]\right)\epsilon~,\\
 \label{condi2}
 &&0=\left[{\cal D}_\mu~,~{\cal D}_D\right]\epsilon=
 {1\over 2}{\widehat \Gamma}_\mu \left(W^\prime\exp(A)+
 \left[W^2\exp(2A)-A^{\prime\prime}\right]{\widetilde \Gamma}_D\right)
 \epsilon~.
\end{eqnarray}
Since in the solitonic brane world
solution $A^\prime$ is discontinuous, to satisfy the last condition $W$
must be discontinuous as well. Then only 1/2 of supersymmetries can be 
preserved, and the corresponding Killing spinor has a definite helicity
w.r.t. ${\widetilde\Gamma}_D$:
\begin{equation}
 {\widetilde\Gamma}_D\epsilon=\eta\epsilon~,
\end{equation}
where $\eta$ is either $+1$ or $-1$. It is then not difficult to see that,
to have a non-trivial solution to ({\ref{Killing2}) compatible with 
the condition (\ref{condi1}), the cosmological constant on the brane must
be vanishing, and we have the following BPS equation:
\begin{equation}
 A^\prime=\eta W\exp(A)~.
\end{equation}
This equation together with the solution (\ref{solutionGB}) then implies
that
\begin{equation}
 W=-\eta \Delta^{-1} {\rm sign}(z)~.
\end{equation}
The Killing spinor is then given by
\begin{equation}
 \epsilon=\exp\left[{1\over 2}A\right]\epsilon_0~,
\end{equation}  
where $\epsilon_0$ is a {\em constant} spinor with helicity $\eta$:
\begin{equation}
 {\widetilde\Gamma}_D\epsilon_0=\eta\epsilon_0~.
\end{equation}
Thus, as we see, 
the flat solution is a BPS solution preserving 1/2 of supersymmetries.

{}The BPS solitonic brane world solution is now stable against
quantum corrections on the brane - 
the cosmological constant on the brane is vanishing as long as
supersymmetry is unbroken.

\section{Solitonic Brane World as a Limit of a Smooth Domain Wall}

{}In the solitonic brane world solution we discussed in the previous
sections the brane is $\delta$-function like. It might appear strange that
such a brane is a soliton. In this section, however, we show that our
solitonic brane can be obtained as a certain limit of a smooth domain wall
solution. Here we would like to emphasize that this is possible as our
solitonic brane world solution does not break diffeomorphisms explicitly
but spontaneously\footnote{This is an important point. Thus, note that if
we introduce a $\delta$-function-like brane source by hand as, say, in 
\cite{RS}, some diffeomorphisms are {\em explicitly} broken. Such
non-solitonic brane then cannot be thought of as a limit of a smooth domain
wall solution as the latter breaks diffeomorphisms spontaneously \cite{COSM}.
Indeed, diffeomorphisms are a local symmetry in this context, so there is a
discontinuity between spontaneously {\em vs.} explicitly broken
diffeomorphisms just as there is a discontinuity between, say, 
spontaneously {\em vs.} explicitly broken $U(1)$ gauge symmetry. 
This as well as other related issues will be
discussed in more detail in \cite{Lang}.}.

\subsection{Setup}

{}In this subsection we discuss the setup within which we will 
construct a smooth solution whose limit gives the aforementioned
solitonic brane world. In parts of the remainder of this section we 
closely follow discussion in \cite{olindo} (also see \cite{Zee}).
Thus, consider a single real scalar field $\phi$
coupled to gravity with the following action\footnote{Here we focus on
the case with one scalar field for the sake of simplicity. In particular,
in this case we can absorb a (non-singular) metric $Z(\phi)$ in
the $(\nabla\phi)^2$ term by a non-linear field redefinition. This cannot
generically be done in the case of multiple scalar fields $\phi^i$, where
one must therefore also consider the metric $Z_{ij}(\phi)$.}:
\begin{eqnarray}
 &&S=M_P^{D-2}\int d^Dx \sqrt{-G}\biggl[ R+
 \lambda\left(R^2-4R^{MN}R_{MN}+R^{MNRS}R_{MNRS}
 \right)\nonumber\\
 &&-{4\over D-2}(\nabla \phi)^2 -V(\phi)\biggr]~,
 \label{actionphi}
\end{eqnarray} 
where $V(\phi)$ is the scalar potential for $\phi$. 
The equations of motion read:
\begin{eqnarray}
 && {8\over{D-2}}\nabla^2\phi=V_\phi~,\\
 \label{einstein1}
 &&R_{MN}-{1\over 2}G_{MN} R
 -{1\over 2}\lambda G_{MN}\left(R^2-4R^{MN}R_{MN}+R^{MNRS}R_{MNRS}
 \right)+\nonumber\\
 &&2\lambda\left(R R_{MN}-2R_{MS}{R^S}_N+R_{MRST}{R_N}^{RST}-
 2R^{RS}R_{MRNS}\right)
 \nonumber\\
 &&={4\over {D-2}}\left[\nabla_M\phi\nabla_N\phi
 -{1\over 2}G_{MN}(\nabla \phi)^2\right]-{1\over 2}G_{MN} V~.
\end{eqnarray}
The subscript $\phi$ in $V_\phi$ denotes derivative w.r.t. $\phi$.

{}In the following we will be interested in solutions to the above equations
of motion where the metric has the form
\begin{equation}\label{warpedy}
 ds^2=\exp(2A)\eta_{\mu\nu}dx^\mu dx^\nu +dy^2~,
\end{equation}
and the warp factor $A$ 
and the scalar field $\phi$ are non-trivial functions
of $y$ but are independent of the $x^\mu$ coordinates.
Here for convenience reasons we choose to work in the coordinate system
$(x^\mu,y)$ instead of $(x^\mu,z)$ as in (\ref{warped}). The two coordinate
systems are related via
\begin{equation}\label{map}
 dy=\exp(A)dz~.
\end{equation}
With this ans{\"a}tz we have the following equations of motion   
for $\phi$ and $A$: 
\begin{eqnarray}
 \label{phi''1}
 &&{8\over{D-2}}\left[\phi_{yy}+(D-1)A_y\phi_y\right]-V_\phi=0~,\\
 \label{phi'A'1}
 &&(D-1)(D-2)(A_y)^2\left[1-(D-3)(D-4)\lambda (A_y)^2\right]
 -{4\over D-2}(\phi_y)^2+V=0~,\\
 \label{A''1}
 &&(D-2)A_{yy}\left[1-2(D-3)(D-4)\lambda(A_y)^2\right]
 +{4\over D-2}(\phi_y)^2=0~,
\end{eqnarray}
where a subscript $y$ denotes derivative w.r.t. $y$.
We can rewrite these equations in terms of
the following first order equations
\begin{eqnarray}\label{BPS1}
 &&\phi_y=\alpha w_\phi\left(1-\lambda\kappa w^2\right)~,\\
 \label{BPS2}
 &&A_y=\beta w~,
\end{eqnarray}
where 
\begin{eqnarray}
 &&\alpha\equiv\sigma {\sqrt{D-2}\over 2}~,\\
 &&\beta\equiv-\sigma {2\over (D-2)^{3/2}}~,\\
 &&\kappa\equiv 2(D-3)(D-4)\beta^2~,
\end{eqnarray}
and $\sigma=\pm 1$. Moreover, the scalar potential $V$ is related
to the function $w=w(\phi)$ via
\begin{equation}
\label{potential}
 V=\left[w_\phi^2 +\Omega\right]\left(1-\lambda\kappa w^2\right)^2- \Omega~,
\end{equation}
where
\begin{equation}
 \Omega\equiv {(D-1)(D-2)\over 4\lambda (D-3)(D-4)}~.
\end{equation}
Note that for $\lambda>0$ the potential (\ref{potential}) is bounded 
below \cite{Zee}.  
Also note that in the $\lambda\rightarrow 0$ limit from (\ref{potential}) 
we recover the familiar expression $V=w_\phi^2-\gamma^2 w^2$, where
$\gamma^2\equiv 4(D-1)/(D-2)^2$.

\subsection{A Domain Wall Solution and the Limit} 

{}Next, we would like to give a simple 
example of a domain wall solution to the
above equations of motion. Thus, let us assume that 
$\lambda>0$, and let 
\begin{equation}
 w=\zeta\phi~.
\end{equation}
The domain wall solution is then given by:
\begin{eqnarray}\label{smooth1}
 &&\phi(y)={1\over{\zeta\sqrt{\lambda\kappa}}}\tanh\left[\alpha\zeta^2
 \sqrt{\lambda\kappa}(y-y_0)\right]~,\\
 \label{smooth2}
 &&A(y)={\beta\over{\alpha\zeta^2\lambda\kappa}}\ln\left(
 \cosh\left[\alpha\zeta^2
 \sqrt{\lambda\kappa}(y-y_0)\right]\right)+A_0~,
\end{eqnarray}
where $y_0$ and $A_0$ are integration constants, which we will set to
zero in the following. Then the point $y=0$ corresponds to the ``center''
of the domain wall, and at this point the warp factor vanishes: $A(0)=0$.

{}Note that in this solution the volume of the $y$ direction, which is
given by
\begin{equation}
 v=\int dy \exp[(D-1)A]~,
\end{equation}
is finite. This implies that gravity is localized on the domain wall.

{}Next, consider the limit where the parameter $\zeta\rightarrow \infty$. 
In this limit the scalar $\phi(y)$ vanishes everywhere, while the warp
factor $A(y)$ is given by:
\begin{equation}
 A(y)=-|y|/\Delta~,
\end{equation} 
where $\Delta$ is given by (\ref{Delta}). If we rewrite this solution
in terms of the $z$ coordinate, we will obtain (\ref{solutionGB}).
Thus, as we see, the solitonic brane world solution discussed in the
previous sections is indeed recovered in the $\zeta\rightarrow\infty$
limit of the above smooth domain wall
solution. Moreover, the smooth solution
interpolates between two AdS minima $\phi_\pm$ 
of the scalar potential $V(\phi)$, and in these minima we have
\begin{equation}
 V(\phi_\pm)=-\Omega~.
\end{equation} 
Note that this is precisely the ``fine-tuned'' bulk vacuum energy density
(\ref{fine}) for which the solitonic brane world solution exists.
In the context of smooth domain wall solutions, however, this does not
appear to be a fine-tuning in the following sense.
Smooth domain walls interpolate between the points $\phi_\pm$ for which the 
r.h.s. of (\ref{BPS1}) is vanishing, that is
\begin{equation}
 w_\phi\left[1-\lambda \kappa w^2\right]\Big|_{\phi_\pm}=0~. 
\end{equation}
Note that {\em a priori} the function
$w(\phi)$ need not have an extremum. Let us assume
that this is indeed the case. Then for 
$\lambda>0$ {\em flat} domain wall solutions exist if and only if the equation
\begin{equation}\label{CONDITION}
 1-\lambda \kappa w^2=0
\end{equation}
has at least two distinct roots (the domain wall then interpolates
between a pair of adjacent points $\phi_\pm$ for which (\ref{CONDITION}) is
satisfied). Note that our ability to rewrite the equations of motion in
terms of the first order equations (\ref{BPS1}) and (\ref{BPS2}) is due to the
fact that we are considering solutions where we have $(D-1)$-dimensional 
Poincar{\'e} invariance on the domain wall. We therefore conclude that in
the context of considering the solitonic brane world solution
(\ref{solutionGB}) as a limit of a smooth domain wall solution the 
{\em classical}\,\footnote{Recall that this condition is stable against 
loop corrections in the context of solitonic brane world with completely
localized gravity.} ``fine-tuning'' condition (\ref{fine}) is not any
less generic than the fine-tuning of the brane cosmological constant.
The latter fine-tuning is absent in the supersymmetric context, so in the
above sense the question whether (\ref{fine}) is a classical fine-tuning
is reduced to the question whether there is any reason why the function
$w(\phi)$ should have (multiple) extrema.

{}Before we end this subsection, let us make the following remarks. In the
supersymmetric context the above smooth domain wall can be shown to
be a BPS state which preserves 1/2 of the original supersymmetries.
The function $w$ in this case plays the role of the superpotential, while
the first order equations (\ref{BPS1}) and (\ref{BPS2}) are the BPS
equations. Also, in the $\zeta\rightarrow\infty$ limit we have
\begin{equation}
 \eta W=\beta w=-{1\over \Delta} {\rm sign}(z)~,
\end{equation}
where $W$ was introduced in subsection B of the previous section.

{}Here we should point out that {\em a priori} there is no guarantee that
the aforementioned smooth domain wall solutions should be embeddable in the
supersymmetric (that is, supergravity) context for, say, $D>4$. In
particular, without the Gauss-Bonnet term such an embedding appears
to be non-trivial \cite{zhukov,gibbons}. Note, however, that the presence
of higher curvature terms might have important implications for this issue.
Thus, for instance, in the above context the superpotential $w(\phi)$ 
need not have extrema for the domain wall to exist, while without the
Gauss-Bonnet term it must. At any rate, whether or not smooth domain walls
of the above type can be embedded in supergravity is outside of the scope
of this paper. Note, however, that the solitonic brane world solution
discussed in the previous sections is clearly embeddable in the
supergravity framework.

\subsection{Normalizable Modes}

{}Let us now study gravity in the above smooth domain wall background. This 
will help us better understand how come gravity is completely localized in the 
solitonic brane world solution discussed above. 
Thus, let us consider small fluctuations around the domain wall solution
(\ref{smooth1}) and (\ref{smooth2}). Here we are going to be interested in
understanding the normalizable modes, so we will not need to include
a source term. The linearized equations of motion then read:
\begin{eqnarray}\label{EOM1oGBsm}
 &&\left[1-2(D-3)(D-4)\lambda(A^\prime)^2\exp(-2A)\right]\Big(
 \partial_\sigma\partial^\sigma 
 H_{\mu\nu} +\partial_\mu\partial_\nu
 H-\partial_\mu \partial^\sigma H_{\sigma\nu}-
 \partial_\nu \partial^\sigma H_{\sigma\mu}-\nonumber\\
 &&\eta_{\mu\nu}
 \left[\partial_\sigma\partial^\sigma H-\partial^\sigma\partial^\rho
 H_{\sigma\rho}\right]+
 H_{\mu\nu}^{\prime\prime}-\eta_{\mu\nu}H^{\prime\prime}+
 (D-2)A^\prime\left[H_{\mu\nu}^\prime-\eta_{\mu\nu}H^\prime\right]-
 \nonumber\\
 &&\left\{\partial_\mu A_\nu^\prime + \partial_\nu A_\mu^\prime -
 2\eta_{\mu\nu}\partial^\sigma A_\sigma^\prime+ (D-2)A^\prime
 \left[\partial_\mu A_\nu +\partial_\nu A_\mu
 -2\eta_{\mu\nu}\partial^\sigma
 A_\sigma\right]\right\}\nonumber+\\
 &&\left\{\partial_\mu\partial_\nu\rho-\eta_{\mu\nu}
 \partial_\sigma\partial^\sigma 
 \rho+\eta_{\mu\nu}\left[(D-2)A^\prime\rho^\prime
 +(D-1)(D-2)(A^\prime)^2\rho\right]\right\}\Big)-\nonumber\\
 &&4(D-4)\lambda\left[A^{\prime\prime}-(A^\prime)^2\right]\exp(-2A)
 \Big(\partial_\sigma\partial^\sigma 
 H_{\mu\nu} +\partial_\mu\partial_\nu
 H-\partial_\mu \partial^\sigma H_{\sigma\nu}-
 \partial_\nu \partial^\sigma H_{\sigma\mu}-\nonumber\\
 &&\eta_{\mu\nu}
 \left[\partial_\sigma\partial^\sigma H-\partial^\sigma\partial^\rho
 H_{\sigma\rho}\right]+
 (D-3)A^\prime\left\{H_{\mu\nu}^\prime-\partial_\mu A_\nu-
 \partial_\nu A_\mu -\eta_{\mu\nu}\left[H^\prime-2\partial^\sigma A_\sigma
 \right]\right\} \Big)+\nonumber\\
 && \eta_{\mu\nu}\rho\left(2(D-2)\left[A^{\prime\prime}-(A^\prime)^2\right]
 \left[1-4(D-3)(D-4)\lambda(A^\prime)^2\exp(-2A)\right]+{4\over{D-2}}
 (\phi^\prime)^2\right)=\nonumber\\
 &&{8\over{D-2}}\eta_{\mu\nu}\phi^\prime\varphi^\prime+\eta_{\mu\nu}
 \varphi V_\phi\exp(2A)~,\\ 
 \label{EOM2oGBsm} 
 &&\left[1-2(D-3)(D-4)\lambda(A^\prime)^2\exp(-2A)\right]\Big(
 \left[\partial^\mu H_{\mu\nu}-\partial_\nu H\right]^\prime 
 -\partial^\mu F_{\mu\nu}+\nonumber\\
 && (D-2)A^\prime \partial_\nu\rho\Big)={8\over{D-2}}\phi^\prime\partial_\nu
 \varphi~,\\
 \label{EOM3oGBsm}
 &&\left[1-2(D-3)(D-4)\lambda(A^\prime)^2\exp(-2A)\right]
 \Big(-\left[\partial^\mu\partial^\nu H_{\mu\nu}-\partial^\mu\partial_\mu H
 \right] +\nonumber\\
 &&(D-2) A^\prime \left[H^\prime-2\partial^\sigma A_\sigma\right]
 -(D-1)(D-2)(A^\prime)^2\rho\Big)+{4\over{D-2}}(\phi^\prime)^2\rho=\nonumber\\
 &&{8\over{D-2}}\phi^\prime\varphi^\prime-\varphi V_\phi\exp(2A)~,\\
 \label{EOM4oGBsm}
 &&\partial_\mu\partial^\mu\varphi+\varphi^{\prime\prime}+(D-2)A^\prime
 \varphi^\prime-{{D-2}\over 8}\varphi V_{\phi\phi}\exp(2A)-\nonumber\\
 &&{1\over 2}\phi^\prime\left[2\partial^\mu A_\mu+\rho^\prime-
 H^\prime\right]-{{D-2}\over 8}\rho V_\phi\exp(2A)=0~,
\end{eqnarray}
where $\varphi$ is the fluctuation for the scalar field $\phi$, and we are 
working in the $(x^\mu,z)$ coordinate system.

{}Since the domain wall breaks diffeomorphisms only spontaneously, we can use
the full $D$-dimensional diffeomorphisms (\ref{diff1}), (\ref{diff2}) and
(\ref{diff3}) to simplify these equations of motion. In fact, such a 
simplification indeed takes place if we perform the gauge transformation with
${\widetilde\xi}_\mu\equiv 0$ and (note that $\phi^\prime$ 
is non-vanishing everywhere) \cite{COSM,zuraCBW}
\begin{equation}
 \omega=-\varphi/\phi^\prime~.
\end{equation}  
It is not difficult to check that this gauge transformation simply removes
$\varphi$ from the above equations of motion. Indeed, under the 
diffeomorphisms we have
\begin{equation}
 \delta \varphi=\phi^\prime\omega~.
\end{equation}
That is, $\varphi$ is {\em not}
a propagating degree of freedom in this gauge \cite{COSM,zuraCBW}.
Note that this uses up some
diffeomorphisms, but the residual diffeomorphisms are sufficient to also gauge
$A_\mu$ away. Indeed, after we remove $\varphi$ from the equations of motion,
we can use the diffeomorphisms with $\omega\equiv 0$ but non-trivial
${\widetilde\xi}_\mu$ to set $A_\mu$ to zero without otherwise changing the
form of the equations of motion. We, therefore, obtain:  
\begin{eqnarray}\label{EOM1oGBsmX}
 &&\left[1-2(D-3)(D-4)\lambda(A^\prime)^2\exp(-2A)\right]\Big(
 \partial_\sigma\partial^\sigma 
 H_{\mu\nu} +\partial_\mu\partial_\nu
 H-\partial_\mu \partial^\sigma H_{\sigma\nu}-
 \partial_\nu \partial^\sigma H_{\sigma\mu}-\nonumber\\
 &&\eta_{\mu\nu}
 \left[\partial_\sigma\partial^\sigma H-\partial^\sigma\partial^\rho
 H_{\sigma\rho}\right]+
 H_{\mu\nu}^{\prime\prime}-\eta_{\mu\nu}H^{\prime\prime}+
 (D-2)A^\prime\left[H_{\mu\nu}^\prime-\eta_{\mu\nu}H^\prime\right]+
 \nonumber\\
 &&\left\{\partial_\mu\partial_\nu\rho-\eta_{\mu\nu}
 \partial_\sigma\partial^\sigma 
 \rho+\eta_{\mu\nu}\left[(D-2)A^\prime\rho^\prime
 +(D-1)(D-2)(A^\prime)^2\rho\right]\right\}\Big)-\nonumber\\
 &&4(D-4)\lambda\left[A^{\prime\prime}-(A^\prime)^2\right]\exp(-2A)
 \Big(\partial_\sigma\partial^\sigma 
 H_{\mu\nu} +\partial_\mu\partial_\nu
 H-\partial_\mu \partial^\sigma H_{\sigma\nu}-
 \partial_\nu \partial^\sigma H_{\sigma\mu}-\nonumber\\
 &&\eta_{\mu\nu}
 \left[\partial_\sigma\partial^\sigma H-\partial^\sigma\partial^\rho
 H_{\sigma\rho}\right]+
 (D-3)A^\prime\left\{H_{\mu\nu}^\prime-
 \eta_{\mu\nu}H^\prime\right\} \Big)+\nonumber\\
 && \eta_{\mu\nu}\rho\Big(2(D-2)\left[A^{\prime\prime}-(A^\prime)^2\right]
 \left[1-4(D-3)(D-4)\lambda(A^\prime)^2\exp(-2A)\right]+\nonumber\\
 &&{4\over{D-2}}
 (\phi^\prime)^2\Big)=0~,\\ 
 \label{EOM2oGBsmX} 
 &&\left[1-2(D-3)(D-4)\lambda(A^\prime)^2\exp(-2A)\right]\Big(
 \left[\partial^\mu H_{\mu\nu}-\partial_\nu H\right]^\prime 
 +(D-2)A^\prime \partial_\nu\rho\Big)=0~,\\
 \label{EOM3oGBsmX}
 &&\left[1-2(D-3)(D-4)\lambda(A^\prime)^2\exp(-2A)\right]
 \Big(-\left[\partial^\mu\partial^\nu H_{\mu\nu}-\partial^\mu\partial_\mu H
 \right] +(D-2) A^\prime H^\prime-\nonumber\\
 &&(D-1)(D-2)(A^\prime)^2\rho\Big)+{4\over{D-2}}(\phi^\prime)^2\rho=0~,\\
 \label{EOM4oGBsmX}
 &&\phi^\prime\left[\rho^\prime-
 H^\prime\right]+{{D-2}\over 4}\rho V_\phi\exp(2A)=0~.
\end{eqnarray}
Here we note that the graviscalar component cannot be gauged away 
after we perform the above gauge fixing.

{}The above equations further simplify in the smooth domain wall background 
discussed in the previous subsection. Thus, let
\begin{equation}
 \chi(z)\equiv\alpha\zeta^2\sqrt{\lambda\kappa} ~y(z)~,
\end{equation}
where $y(z)$ can be computed using the map (\ref{map}). We have:
\begin{eqnarray}\label{EOM1oGBsmXYZ}
 &&\left[1+{4(D-4)\over{D-2}}\lambda\zeta^2\right]
 \left(\partial_\sigma\partial^\sigma 
 H_{\mu\nu} +\partial_\mu\partial_\nu
 H-\partial_\mu \partial^\sigma H_{\sigma\nu}-
 \partial_\nu \partial^\sigma H_{\sigma\mu}-
 \eta_{\mu\nu}
 \left[\partial_\sigma\partial^\sigma H-\partial^\sigma\partial^\rho
 H_{\sigma\rho}\right]\right)+\nonumber\\
 &&H_{\mu\nu}^{\prime\prime}-\eta_{\mu\nu}H^{\prime\prime}+
 (D-2)A^\prime\left[1+{4(D-3)(D-4)\over(D-2)^2}\lambda\zeta^2\right]
 \left[H_{\mu\nu}^\prime-\eta_{\mu\nu}H^\prime\right]+\nonumber\\
 &&\partial_\mu\partial_\nu\rho-\eta_{\mu\nu}
 \partial_\sigma\partial^\sigma 
 \rho+\eta_{\mu\nu}(D-2)A^\prime\rho^\prime+\nonumber\\
 &&\eta_{\mu\nu}\rho\exp(2A)\left\{\zeta^2\left[2-3\cosh^{-2}(\chi)\right]+
 2\Omega\tanh^2(\chi)\right\}=0~,\\
 \label{EOM2oGBsmXYZ} 
 &&
 \left[\partial^\mu H_{\mu\nu}-\partial_\nu H\right]^\prime 
 +(D-2)A^\prime \partial_\nu\rho=0~,\\
 \label{EOM3oGBsmXYZ}
 &&
 -\left[\partial^\mu\partial^\nu H_{\mu\nu}-\partial^\mu\partial_\mu H
 \right] +(D-2) A^\prime H^\prime+
 \rho\exp(2A) 
 \left[\zeta^2 \cosh^{-2}(\chi)-2\Omega\tanh^2(\chi)\right]=0~,\\
 \label{EOM4oGBsmXYZ}
 &&\rho^\prime-
 H^\prime +{8(D-3)(D-4)\over {D-2}}\lambda\left[\zeta^2+\Omega\right]
 A^\prime \rho =0~.
\end{eqnarray}
Here we note that not all of these equations are independent.
Thus, differentiating (\ref{EOM1oGBsmXYZ}) with $\partial^\mu$, we obtain  
an equation which is identically satisfied once we take into account
(\ref{EOM2oGBsmXYZ}) as well as the on-shell expression for $A$.
Also, if we take the trace of (\ref{EOM1oGBsmXYZ}), then we obtain an equation
which is identically satisfied once we take into account (\ref{EOM2oGBsmXYZ}),
(\ref{EOM3oGBsmXYZ}) and (\ref{EOM4oGBsmXYZ}) as well as the on-shell 
expression for $A$. This, as usual, is a consequence of Bianchi identities.

{}Now we are ready to discuss normalizable modes in the above domain wall
background. Let us first consider the normalizable modes for the graviscalar
$\rho$. Thus, we can eliminate $H_{\mu\nu}$ from (\ref{EOM2oGBsmXYZ}),
(\ref{EOM3oGBsmXYZ}) and (\ref{EOM4oGBsmXYZ}), which gives us the following
second order equation for $\rho$:
\begin{equation}\label{moderho}
 \rho^{\prime\prime}+\psi A^\prime\rho^\prime+\partial^\mu\partial_\mu\rho
 +F\rho=0~,
\end{equation}  
where
\begin{equation}
 \psi\equiv D+{8(D-3)(D-4)\over{D-2}}\lambda\zeta^2~,
\end{equation}
and
\begin{equation}
 F(z)\equiv 2(A^\prime)^2\left[1+{2(D-3)(D-4)\over{D-2}}\lambda\zeta^2\right]
 +2 A^{\prime\prime}\left[D-2+{6(D-3)(D-4)\over{D-2}}\lambda\zeta^2\right]~.
\end{equation}
Let us now assume that $\rho$ satisfies the $(D-1)$-dimensional Klein-Gordon 
equation
\begin{equation}
 \partial^\mu\partial_\mu\rho=m^2\rho~.
\end{equation}
In the following we will assume that $m^2\geq 0$. As to the $m^2<0$ modes,
they cannot be normalizable - indeed, the domain wall is a kink-like 
object, and is therefore stable, so no tachyonic modes are normalizable.

{}We need to understand the asymptotic behavior of $\rho$ at large $z$.
To do this, it is convenient to rescale $\rho$ as follows:
\begin{equation}
 \rho\equiv{\widetilde \rho}\exp\left[-{1\over 2}\psi A\right]~.
\end{equation}
The equation (\ref{moderho}) then reads:
\begin{equation}
 {\widetilde \rho}^{\prime\prime}+\left[m^2+F-{1\over 2}\psi
 A^{\prime\prime}-{1\over 4}\psi^2(A^\prime)^2\right]{\widetilde \rho}=0~.
\end{equation}
Note that at large $z$ the functions $F$, $A^{\prime\prime}$ and
$(A^\prime)^2$ go to zero as $\sim 1/z^2$. We therefore have the following
leading behavior for ${\widetilde \rho}$ at large $z$:
\begin{equation}
 {\widetilde \rho}(z)=C_1 \cos(mz)+C_2 \sin(mz)~,
\end{equation}
where $C_1,C_2$ are some constant coefficients.

{}Next, note that the norm for the graviscalar is given by
\begin{equation}
 ||\rho||^2\propto\int dz~\exp(DA)\rho^2~,
\end{equation}
where
the measure $\exp(DA)$ comes from $\sqrt{-G}$. In terms of ${\widetilde
\rho}$ we have
\begin{equation}
 ||\rho||^2\propto\int dz~\exp\left[-{8(D-3)(D-4)\over{D-2}}\lambda\zeta^2 
 A\right]
 {\widetilde \rho}^2~.
\end{equation}
Since $A$ goes to $-\infty$ at large $z$, we conclude that
none of the $m^2>0$ modes
are even plane-wave normalizable. Moreover, since the 
function $F$ in (\ref{moderho}) is non-trivial, we do not have a quadratically
normalizable zero mode either. Thus, we conclude that $\rho$ is {\em not} a
propagating degree of freedom in the above background, and should be set to
zero.

{}Next, let us turn to the normalizable modes for the graviton 
$H_{\mu\nu}$. From (\ref{EOM2oGBsmXYZ}),
(\ref{EOM3oGBsmXYZ}) and (\ref{EOM4oGBsmXYZ}) it follows that, since $\rho
\equiv 0$, we have
\begin{equation}
 \partial^\mu H_{\mu\nu}^\prime=H^\prime=0~.
\end{equation}
This then implies that we can use the residual $(D-1)$-dimensional
diffeomorphisms (for which $\omega\equiv 0$, and ${\widetilde\xi}_\mu$ are 
independent of $z$) to bring $H_{\mu\nu}$ into the transverse-traceless
form:
\begin{equation}
 \partial^\mu H_{\mu\nu}=H=0~.
\end{equation}
It then follows from (\ref{EOM1oGBsmXYZ}) that for the modes of the form 
\begin{equation}
 H_{\mu\nu}=\xi_{\mu\nu}(x^\rho) \Sigma(z)~,
\end{equation}
where
\begin{equation}
 \partial^\sigma\partial_\sigma \xi_{\mu\nu}=m^2\xi_{\mu\nu}~,
\end{equation}
the $z$-dependent part of $H_{\mu\nu}$ satisfies the following equation:
\begin{equation}\label{modeH}
 \Sigma^{\prime\prime} +\Psi_1 A^\prime\Sigma^\prime +\Psi_2^2 m^2 
 \Sigma=0~,
\end{equation}
where
\begin{eqnarray}
 &&\Psi_1\equiv (D-2)\left[1+{4(D-3)(D-4)\over(D-2)^2}\lambda\zeta^2\right]~,\\
 &&\Psi_2\equiv \left[1+{4(D-4)\over{D-2}}\lambda\zeta^2\right]^{1\over 2}~. 
\end{eqnarray}
Let us rescale $\Sigma$ as follows:
\begin{equation}
 \Sigma\equiv{\widetilde\Sigma}\exp\left[-{1\over 2}\Psi_1 A\right]~.
\end{equation}
The equation (\ref{modeH}) now reads:
\begin{equation}
 {\widetilde\Sigma}^{\prime\prime}+\left[\Psi_2^2 m^2 -{1\over 2}\Psi_1 
 A^{\prime\prime}-{1\over 4}\Psi_1^2 (A^\prime)^2\right]{\widetilde\Sigma}=0~.
\end{equation}
At large $z$ we therefore have:
\begin{equation}
 {\widetilde \Sigma}(z)=D_1 \cos\left(\Psi_2 m z\right) +D_2 
 \sin\left(\Psi_2 m z\right)~,
\end{equation}
where $D_1,D_2$ are some constant coefficients. 

{}Next, note that the norm for the graviton is given by
\begin{equation}
 ||H_{\mu\nu}||^2\propto\int dz~\exp[(D-2)A]\Sigma^2~,
\end{equation}
where, unlike the graviscalar case,
the measure $\exp[(D-2)A]$ comes from $\sqrt{-G}R$. In terms of ${\widetilde
\Sigma}$ we have
\begin{equation}
 ||H_{\mu\nu}||^2
 \propto\int dz~\exp\left[-{4(D-3)(D-4)\over{D-2}}\lambda\zeta^2 
 A\right]
 {\widetilde \Sigma}^2~.
\end{equation}
Thus, we see that 
none of the $m^2>0$ modes
are even plane-wave normalizable. Unlike the graviscalar 
case, however, we do 
have a quadratically normalizable zero mode for $H_{\mu\nu}$.
This zero mode is given by $\Sigma^\prime=0$. 

{}Thus, as we see, in the above
smooth domain wall background the only propagating degree of freedom is the
zero mode of $H_{\mu\nu}$ corresponding to $(D-1)$-dimensional 
gravity. This result then makes it clear why in the $\zeta\rightarrow\infty$
limit, where we recover the aforementioned $\delta$-function-like solitonic
brane world solution, gravity is completely localized on the brane.

\section{Conjecture}

{}In the previous sections we discussed a solitonic brane world solution
with completely localized gravity arising in the
Einstein-Hilbert-Gauss-Bonnet theory with a cosmological term. In $D=5$ 
the Gauss-Bonnet combination is the only higher curvature term that we can 
consider in this context\footnote{Here the following remark is in order.
The background in such warped compactifications is not affected by higher
curvature terms constructed solely 
from the Weyl tensor (which is the ``traceless'' part of the Riemann tensor)
as such terms vanish in conformally flat backgrounds. Note, however, that
such terms would 
affect the graviton propagator in the bulk.}. 
For higher $D$, however, we can add higher Euler invariants
({\em e.g.}, in $D=7$ we can also include the 6-dimensional Euler 
invariant). In these cases we also expect to have similar
solitonic brane world solutions.

{}Regardless of an explicit field theory realization of such a solitonic
brane world, one question that immediately arises is whether we can have 
matter localized on the brane. If this brane is a D-brane-like object, then
we could hope to have gauge and matter fields localized on the brane. In
this context we would like to ask whether we can embed our scenario in the
string theory framework. In fact, here we would like to take this a bit 
further and ask whether (at least in some cases) D-branes can be identified
as such solitonic brane world solutions. In this section we will propose
a conjecture (in a weak as well as strong form) according to which the
answers to the above questions are positive. In the following we will focus 
on the case of 3-branes, although we expect that generalizations to other
branes should also be possible.

\subsection{D3-branes at Conical Singularities} 

{}In this subsection we will review some facts concerning parallel D3-branes 
near a conical singularity. Our discussion here will closely follow 
\cite{KW}\footnote{This setup was originally discussed in \cite{Kehagias}.}. 
Thus, consider Type IIB in the 
presence of (large but finite number) 
$N$ D3-branes at the conical singularity located
at $r=0$ in the 6-dimensional non-compact Ricci-flat manifold $Y_6$ with the
metric 
\begin{equation}
 g_{IJ}dx^I dx^J=(dr)^2+r^2 \gamma_{ij}dx^i dx^j~.
\end{equation}
Here $\gamma_{ij}$ is a metric on a 5-dimensional manifold $X_5$, and the 
$r=0$ point is a singularity unless $X_5$ is a 5-sphere. The condition that 
$Y_6$ is Ricci-flat implies that $X_5$ is an Einstein manifold of positive
curvature. Note that in the special case where $X_5$ is a 5-sphere ${\bf
S}^5$ the manifold $Y_6$ is actually flat.

{}In the above background
the 10-dimensional metric has the following form ($\eta_{\mu\nu}$ is the 
flat 4-dimensional Minkowski metric):
\begin{equation}
 ds^2=Q^{-1/2}(r) \eta_{\mu\nu}dx^\mu dx^\nu +Q^{1/2}(r)
 \left[(dr)^2+r^2 \gamma_{ij}dx^i dx^j\right]~,
\end{equation}
where
\begin{equation}
 Q(r)\equiv 1+{L^4\over r^4}~,~~~L^4\equiv 4\pi g_s N (\alpha^\prime)^2~.
\end{equation}
The near horizon limit of the above geometry, that is, the limit 
$r\rightarrow 0$, coincides with that of AdS$_5\times X_5$. According to
the arguments in \cite{malda}, we expect that the field theory on the
D3-branes at the above conical singularity should be dual to Type IIB
string theory on AdS$_5\times X_5$ (with $N$ units of 5-form flux on $X_5$). 

{}Let us now discuss the field theory on the branes. In the special case
where $X_5={\bf S}^5$ the manifold $Y_6$ is smooth (it is simply 
${\bf R}^6$). The gauge theory on the branes is then ${\cal N}=4$ 
$U(N)$ SYM theory, which is conformal. This fact is consistent with the
conformal property of AdS$_5$ and the aforementioned duality between
the gauge theory on the branes and Type IIB on AdS$_5\times X_5$
\cite{GKP,WITT}.    

{}Let us now consider cases where $X_5$ is {\em not} a 5-sphere (but is
still a compact Einstein manifold). In this case supersymmetry is at least 
partially broken. The simplest examples of such manifolds are orbifolds
of ${\bf S}^5$: $X_5={\bf S}^5/\Gamma$ \cite{KS}, where $\Gamma$ is a finite
discrete subgroup of $SO(6)$, or, more precisely, of ${\rm Spin}(6)$
as we are dealing with a theory containing fermions. Note that 
the latter group is the $R$-parity group of ${\cal N}=4$ SYM.
The gauge theories on branes at the corresponding orbifold singularities
were discussed in detail in \cite{LNV,BKV}\footnote{The generalization to
gauge theories on branes at orientifold singularities was subsequently
discussed in \cite{orient}.}. Thus, if $\Gamma\subset SU(3) (SU(2))$, then
the corresponding gauge theory is ${\cal N}=1$ (${\cal N}=2$)
supersymmetric. Otherwise, supersymmetry is completely broken. 
Here we would like to consider some simple
examples of such theories which capture the main point we would like to
make.

{}Thus, let us consider the simplest example of such a theory. Let
${\Gamma}={\bf Z}_2$, whose generator $R$ has the following action on the
complex coordinates $z_1,z_2,z_3$ on the 5-sphere (the 5-sphere is given by
$|z_1|^2+|z_2|^2+|z_3|^2=\rho^2$, where $\rho$ is its radius): $R
z_{1,2}=-z_{1,2}$, $R z_3=z_3$. The gauge theory on the branes is
given by the ${\cal N}=2$ $U(n)\otimes U(n)$ gauge theory with 2 copies of
hypermultiplets in $({\bf n},{\bf {\overline n}})(+1,-1)$ and    
$({\bf {\overline n}}, {\bf n})(-1,+1)$, where the $U(1)$ charges are given
in parenthesis. Note that nothing is charged under the diagonal $U(1)_S$
(which corresponds to the brane center-of-mass degree of freedom), but
the matter fields are charged under the anti-diagonal $U(1)_A$. (The
generators of these $U(1)$'s are given by $Q_S={1\over\sqrt{2}}(Q_1+Q_2)$
and $Q_A={1\over\sqrt{2}}(Q_1-Q_2)$, where $Q_1,Q_2$ are the generators of
the original $U(1)$'s.) This implies that $U(1)_A$ runs, and decouples in
the infra-red. On the other hand, the non-Abelian part of the gauge group
is conformal.

{}Let us briefly discuss another example of this type, which has ${\cal
N}=1$ supersymmetry. Let the orbifold group be $\Gamma={\bf Z}_3$, whose
generator $\theta$ has the following action: $\theta z_{1,2,3}=
\exp(2\pi i/3)z_{1,2,3}$. Then the gauge theory is the ${\cal N}=1$ 
$U(n)\otimes U(n)\otimes U(n)$ gauge theory with 3 copies of chiral
multiplets in $({\bf n},{\bf {\overline n}},{\bf 1})(+1,-1,0)$,
$({\bf 1}, {\bf n},{\bf {\overline n}})(0,+1,-1)$ and 
$({\bf {\overline n}},{\bf 1},{\bf n})(-1,0,+1,)$, call them
$A_a,B_a,C_a$, $a=1,2,3$, and the renormalizable superpotential
\begin{equation}
 {\cal W}=\epsilon_{abc} A_a B_b C_c~.
\end{equation}
Once again, there is
nothing charged under the diagonal $U(1)$, while the matter fields carry
non-trivial charges under the other two linear combinations. These two
$U(1)$'s decouple in the infra-red, and the non-Abelian part of the gauge
group is conformal. 

{}The point that we would like to make here, however, is that the gauge
theories in the above examples as well as for {\em any} other non-trivial
choice of the orbifold group $\Gamma$ are conformal only in the infra-red.
In particular, in the orbifold examples we always have running $U(1)$'s
under which matter fields are charged. This implies that the brane
world-volume theory is actually not conformal, in particular, in the
ultra-violet we have non-trivial loop corrections.

{}Before we turn to the upshot of the above discussion, let us give one
more example, where $X_5$ is not an orbifold of ${\bf S}^5$. This example is 
instructive as already the non-Abelian part of the corresponding gauge theory
is not conformal in the ultra-violet. Thus, consider the case where $X_5=
T^{1,1}=(SU(2)\times SU(2))/U(1)$ \cite{KW}. 
The gauge theory then is given by
the ${\cal N}=1$ $U(n)\otimes U(n)$ gauge theory with 2 copies of chiral
multiplets in $({\bf n},{\bf {\overline n}})(+1,-1)$ and    
$({\bf {\overline n}}, {\bf n})(-1,+1)$, call them $A_a,B_a$, $a=1,2$, and the
non-renormalizable superpotential
\begin{equation}
 {\cal W}=\epsilon_{ab}\epsilon_{cd} A_a B_c A_b B_d~.
\end{equation}
As in the $X_5={\bf S}^5/{\bf Z}_2$ 
example\footnote{In fact, the $X_5={\bf S}^5/{\bf Z}_2$
and $X_5=T^{1,1}$ cases are related as follows \cite{KW}. 
There is a fixed circle in ${\bf S}^5/{\bf Z}_2$. Blowing up this fixed circle
breaks half of the supersymmetries, and deforms the ${\cal N}=2$
gauge theory corresponding to 
the $X_5={\bf S}^5/{\bf Z}_2$ case to the ${\cal N}=1$ gauge theory
corresponding to the $X_5=T^{1,1}$ case.}, here we also have a running
(anti-diagonal) $U(1)$. Note, however, that already the non-Abelian part of 
the gauge group is {\em not} conformal in the ultra-violet. As to the 
infra-red, the anti-diagonal $U(1)$ decouples (the diagonal $U(1)$ is free to 
begin with), and the non-Abelian part of the theory flows into a conformal
fixed point.

\subsection{Gravity on D-branes}

{}Thus, as we see, as long as $X_5$ is not a 5-sphere, the gauge theory is not
conformal in the ultra-violet, albeit it is conformal in the 
infra-red\footnote{If we include orientifold planes \cite{orient}, then 
one can construct examples where the gauge 
theory (at least in the ${\cal N}=1$ cases) is not conformal even in the 
infra-red. In these cases, however, some caution is needed due to subtleties
discussed in \cite{KaSh,KST}.}. We therefore have non-trivial loop corrections
in the ultra-violet. This has important consequences.
Thus, as was pointed out in \cite{DGP,DG}, if we have non-conformal theory 
on a brane, loop corrections are generically expected to generate graviton
propagator on the brane. That is, at the quantum level we expect 
(among other terms) the
Einstein-Hilbert term to be generated in the world-volume of D3-branes
with non-conformal theories. The question we would like to ask here is whether
we can understand the presence of this term in the world-volume theory of 
D3-branes in the dual Type IIB picture. Here we would like to make a conjecture
which implies that the answer to this question is positive.

\begin{center}
 {\em Weak Form of the Conjecture}
\end{center}

{}From the above discussion it is clear that the gauge theory on D3-branes
at the conical singularity in $Y_6$ can be dual to Type IIB on AdS$_5\times
X_5$ only in the infra-red limit of the gauge theory unless $X_5={\bf S}_5$.
It is precisely in the latter case that $Y_6$ is actually non-singular (and,
moreover, flat), and we have maximal supersymmetry. In other cases 
supersymmetry is at least partially broken. We can therefore expect that the
AdS$_5\times X_5$ background is corrected by higher curvature (or, more 
generally, higher derivative) terms\footnote{As was argued in 
\cite{Banks,Kallosh},
the AdS$_5\times {\bf S}_5$ background is an exact solution of Type IIB, which
is due to the fact that we have maximal supersymmetry in this case.}. 
That is, here we propose that, if $X_5\not={\bf S}_5$, AdS$_5\times X_5$ 
with $N$ units of 5-form flux is {\em not} an exact background of Type IIB.
Moreover, we propose that we have {\em gravity} 
on D3-branes, that is,
we have the Einstein-Hilbert term in their world-volume action. 
In the gauge theory
language this term arises from loop corrections due to non-conformal matter
fields\footnote{In the $X_5={\bf S}^5/\Gamma$ orbifold cases, where 
non-conformality is due to running $U(1)$'s, the Einstein-Hilbert term
is expected to arise at the two-loop order.
In the cases where already the non-Abelian part of the gauge group
is non-conformal, this is expected to occur at the one-loop order.
This can be seen by using arguments similar to those in \cite{BKV}.}. 
Here we conjecture that there is a dual Type IIB description, which is
valid beyond the infra-red limit of the gauge theory, and in this description
the effects corresponding to having gravity in the D-brane world-volume 
are due to higher curvature terms (which are responsible for the fact that
AdS$_5\times X_5$ is not an exact background). Here we note that, if such
a duality indeed holds, the fact that
AdS$_5\times X_5$ cannot be an exact background is evident from the fact that 
the dual gauge theory is not conformal in the ultra-violet. On the other hand,
the origin of the aforementioned higher curvature terms {\em a priori}
might be 
less evident. Here we would like to stress that these are {\em not} the 
higher curvature terms (such as the $({\rm Weyl})^4$ terms)
that are already present in a flat 10-dimensional 
background of Type IIB \cite{Gutperle}.
Rather, these higher curvature terms are intrinsically due to the
compactification. Thus, for instance, let us consider the orbifold cases
$X_5={\bf S}^5/\Gamma$. In these cases we have twisted sector fields 
(including those that are massive) which 
contribute into the higher curvature terms, and these terms are intrinsically
``5-dimensional''. Note that the appearance of these intrinsically 
``5-dimensional'' higher curvature terms goes hand-by-hand with 
(partial) supersymmetry breaking.

\begin{center}
 {\em Strong Form of the Conjecture}
\end{center}

{}The above (form of the)
conjecture is somewhat vague in the sense that it does not
specify how the AdS$_5\times X_5$
background is modified due to the aforementioned higher curvature terms.
Clearly, this is not a simple question, but, nonetheless, here we would like
to make a guess based on the following observations. First,
in the cases where the
D-brane world-volume gauge theory is conformal in the infra-red, we 
expect that the dual Type IIB background should be at least asymptotically
AdS$_5\times X_5$. Second, we have gravity on D-branes. Third, at least in some
limit D-branes are $\delta$-function-like objects. Here we conjecture that
if we take into account higher curvature terms in Type IIB compactification
on $X_5$ with $N$ units of 5-form flux on $X_5$, 
(at least in some cases) in the 5-dimensional
space transverse to $X_5$ we have {\em solitonic} solutions, which (at least
in a certain limit) are $\delta$-function like, and couple to
the Ramond-Ramond 4-form. We propose that these solitonic solutions are 
nothing but D3-branes. Moreover, there is gravity localized on these solitonic
D3-branes, while the 5-dimensional space transverse to $X_5$ is asymptotically
AdS$_5$. 
Since here we are talking about solitonic $\delta$-function-like branes,
in the light of our discussions in the previous sections here we
also propose that
gravity is {\em completely} localized\footnote{The reason why we expect 
gravity to be completely localized on such branes is due to the fact that
the latter are $\delta$-function-like
solitonic solutions which do not break diffeomorphisms 
explicitly but spontaneously. This then implies that the graviscalar is 
a gauge degree of freedom. On the other hand, if gravity is not completely
localized, that is, if we have massive Kaluza-Klein modes propagating in the
bulk as in, say, \cite{RS}, then at short distances (that is, large momenta)
gravity is expected to become 5-dimensional, which requires that the coupling
of the graviscalar to the {\em non-conformal} 
brane matter be non-vanishing \cite{zuraRS}, but this is not possible
if we can gauge the graviscalar away. A more
detailed discussion of this point will be given in \cite{Lang}.} 
on these branes, that is, it does not
propagate in the 5th dimension (albeit there are heavy Kaluza-Klein modes
propagating in the other 5 directions along $X_5$, which 
is compact)\footnote{Here we would like to make the following remark.
In the above picture, in the Type IIB language the analog of the
``fine-tuning'' relation (\ref{fine}) should arise as a result of the fact 
that the relevant higher curvature terms come from the compactification on 
$X_5$, so that the corresponding couplings should be determined by the
geometry of $X_5$. On the other hand, the volume of $X_5$ as well as the bulk
vacuum energy density are expected to be related to the 5-form flux, which
is quantized.}. 

{}One of the implications of this conjecture would be that the   
loop corrections, apart from those due to the heavy KK/string 
thresholds, are 4-dimensional and not 5-dimensional\footnote{The theory is
in some vague sense ``holographic''. At present it is unclear
whether this can in any way be tied to the observations of 
\cite{DGP0,witten,DGP,zura,zuraCBW,DG} 
that bulk supersymmetry might in some cases
control the brane cosmological constant.}, 
just as in the solitonic brane world 
solution we discussed in this paper. Moreover, in this picture D-branes are
{\em non-singular} solitonic solutions of full Type IIB string 
theory\footnote{Note that according to the stronger form of the above 
conjecture, in the corresponding cases we do not really have two (gauge theory
{\em vs.} Type IIB) descriptions. Rather, D-branes are part of the Type IIB
background. Note that $X_5={\bf S}^5$ case (and, perhaps, some other cases)
is exceptional in this sense. Nonetheless, perhaps this case can be thought
of as some limit of the generic situation described in the above conjecture.}.
Finally, this conjecture relates observations of this
paper, the scenarios discussed in \cite{DGP,DG}, and string theory in the 
spirit of \cite{malda}, which might at first seem unrelated. 

\acknowledgments

{}We would like to thank Michael Gutperle, Kentaro Hori, Tom Taylor 
and Cumrun Vafa for
valuable discussions. Parts of this work were completed while Z.K. was
visiting at Northeastern University and 
Harvard University.
This work was supported in part by the National Science Foundation.
Z.K. would also like to thank Albert and Ribena Yu for financial support.

\end{document}